\def\year{2018}\relax
\documentclass[letterpaper]{article} %
\usepackage{aaai18}  %
\usepackage{times}  %
\usepackage{helvet}  %
\usepackage{courier}  %
\usepackage{url}  %
\usepackage{graphicx}  %
\usepackage{latexsym}

\usepackage{booktabs}
\usepackage{array}
\usepackage{multirow}
\usepackage{color}
\usepackage{blindtext}
\usepackage[skip=4pt]{caption}
\usepackage{subcaption}
\usepackage{enumitem}
\usepackage{tikz}
\usepackage{amsmath,bm,amssymb}
\usepackage{comment}
\usepackage{tabulary}
\usepackage{makecell}
\usepackage{tcolorbox}
\usepackage{cleveref}
\usepackage{balance}
\usetikzlibrary{calc,decorations.pathreplacing}

\newcommand{\addFigure}[2]{\includegraphics[width=#1]{figs/#2}}

\newcommand{\hide}[1]{} %

\newcommand{\para}[1]{\vspace{0.01in}\noindent\textbf{#1 }}
\newcommand{\figref}[1]{Fig.~\ref{#1}} %
\newcommand{\tableref}[1]{Table~\ref{#1}} 
\newcommand{\communityname}[1]{{\small\sf #1}\xspace}
\newcommand{\mediancommunitysize}{341\xspace}
\newcommand{\avgparentsize}{180\xspace}

\title{Tracing Community Genealogy: \\ How New Communities Emerge from the Old}
\author{Chenhao Tan\\
Department of Computer Science\\
University of Colorado Boulder\\
Boulder, CO, 80309 \\
{\tt chenhao@chenhaot.com}
}

\newcommand{\citet}[1]
{\citeauthor{#1}~\shortcite{#1}}
\newcommand{\citep}{\cite}

\begin{document}
\maketitle

\begin{abstract}

The process by which  new communities emerge is a central research issue in the social sciences.
While a growing body of research analyzes the formation of a single community by examining social networks between individuals, we introduce a novel community-centered perspective.
We highlight the fact that the context in which a new community emerges 
contains numerous existing communities.
We reveal the emerging process of communities by tracing 
their early members' previous community memberships.

Our testbed is Reddit, a website that consists of tens of thousands of user-created communities.
We analyze a dataset that spans over a decade and includes the 
posting history of users on Reddit from its inception to April 2017.
We first propose a computational framework for 
building genealogy graphs between communities.
We present the first large-scale characterization of such genealogy graphs. 
Surprisingly, basic graph properties, such as the number of parents and max parent weight,
converge quickly despite the fact that the number of communities increases rapidly over time.
Furthermore, we investigate the connection between a community's origin and its future growth.
Our results show that strong parent connections are associated with future community growth, confirming the importance of existing community structures in which a new community emerges.
Finally, we turn to the individual level and examine the characteristics of early members.
We find that a diverse portfolio across existing communities is the most important predictor for becoming an early member in a new community.

\end{abstract}

\section{Introduction}

\begin{figure}[t]
    \centering
    \addFigure{0.32\textwidth}{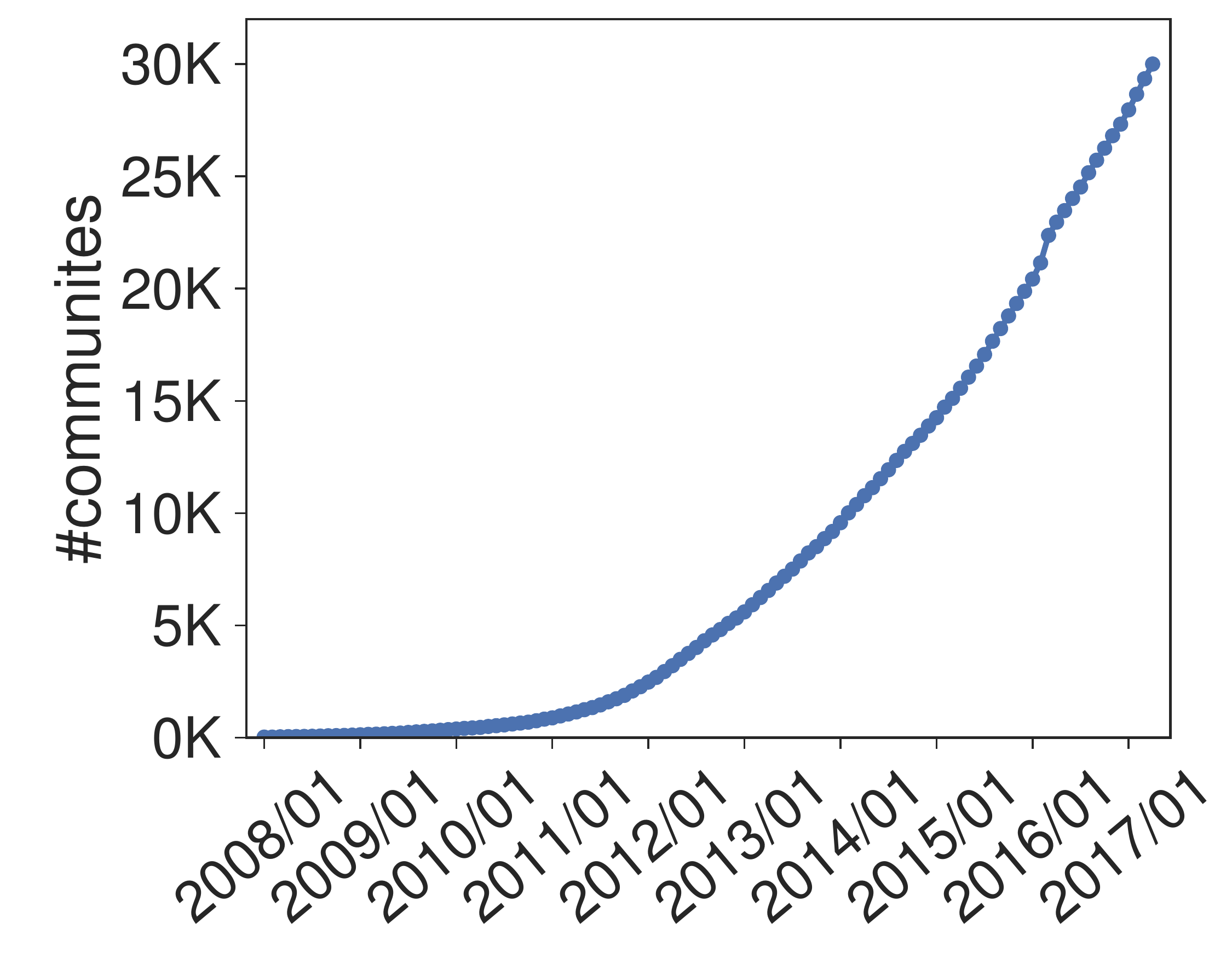}
    \caption{
    The total number of communities with more than 100 members on Reddit by each month: 
    the number of communities has soared since 2008, when 
    users started to be able to create their own communities.
    }
    \label{fig:sub_count}
\end{figure}

\begin{figure*}[t]
    \centering
    \begin{subfigure}[t]{0.26\textwidth}
        \addFigure{\textwidth}{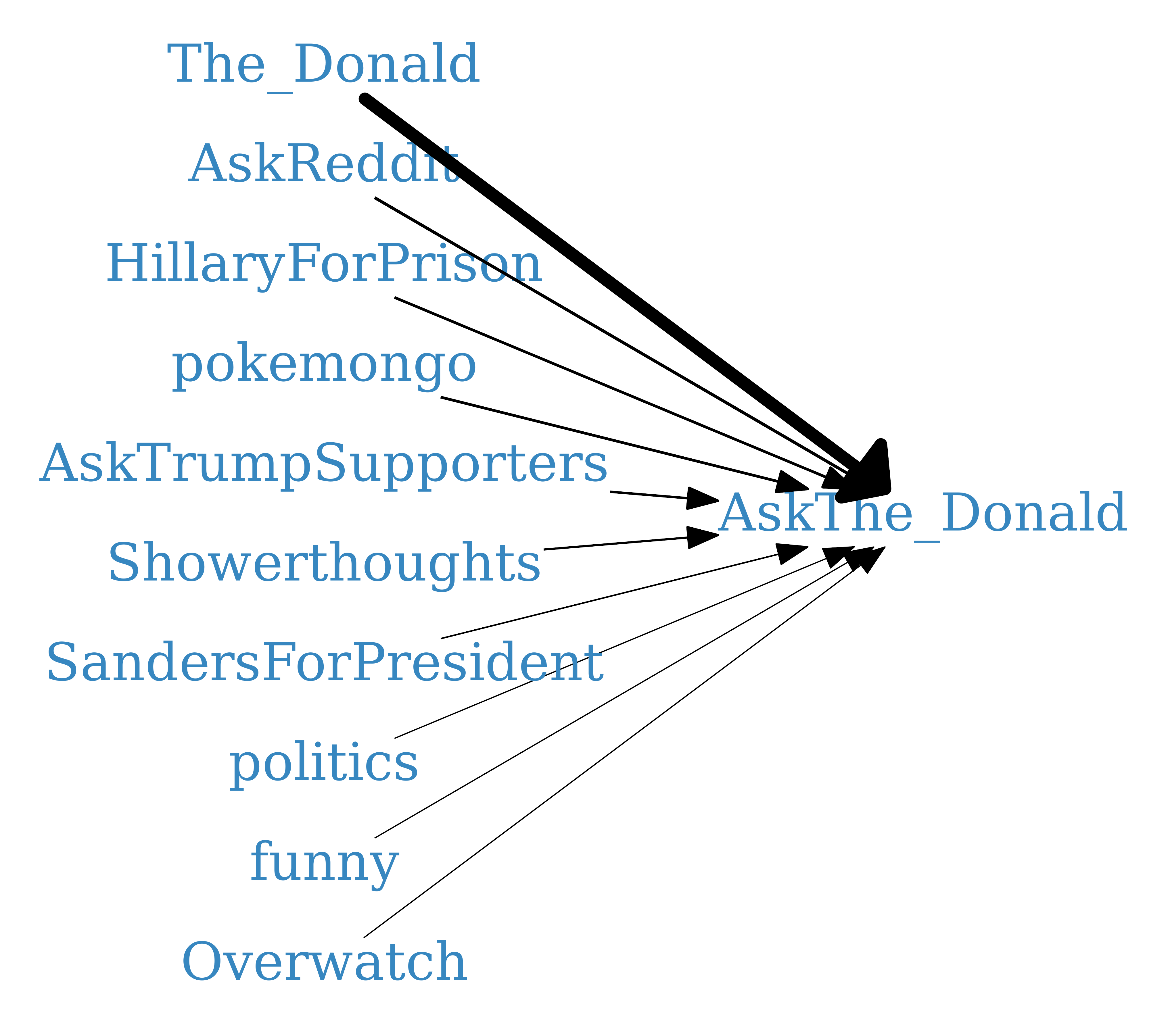}
        \caption{Top parents of \communityname{AskThe\_Donald}}
        \label{fig:intro_p1}
    \end{subfigure}
    \hfill
    \begin{subfigure}[t]{0.26\textwidth}
        \addFigure{\textwidth}{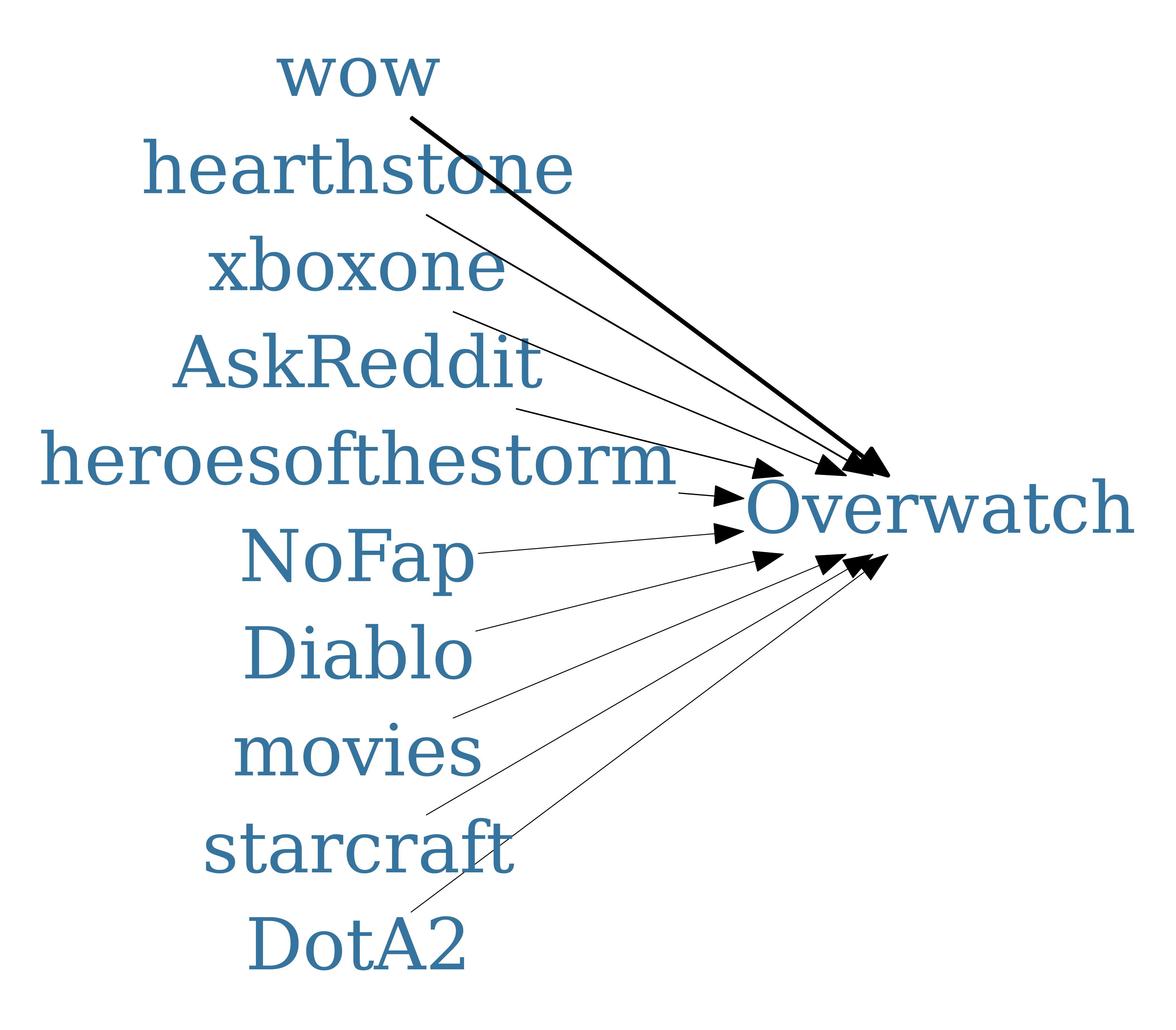}
        \caption{Top parents of \communityname{Overwatch}}
        \label{fig:intro_p2}
    \end{subfigure}
    \hfill
    \begin{subfigure}[t]{0.44\textwidth}
        \addFigure{\textwidth}{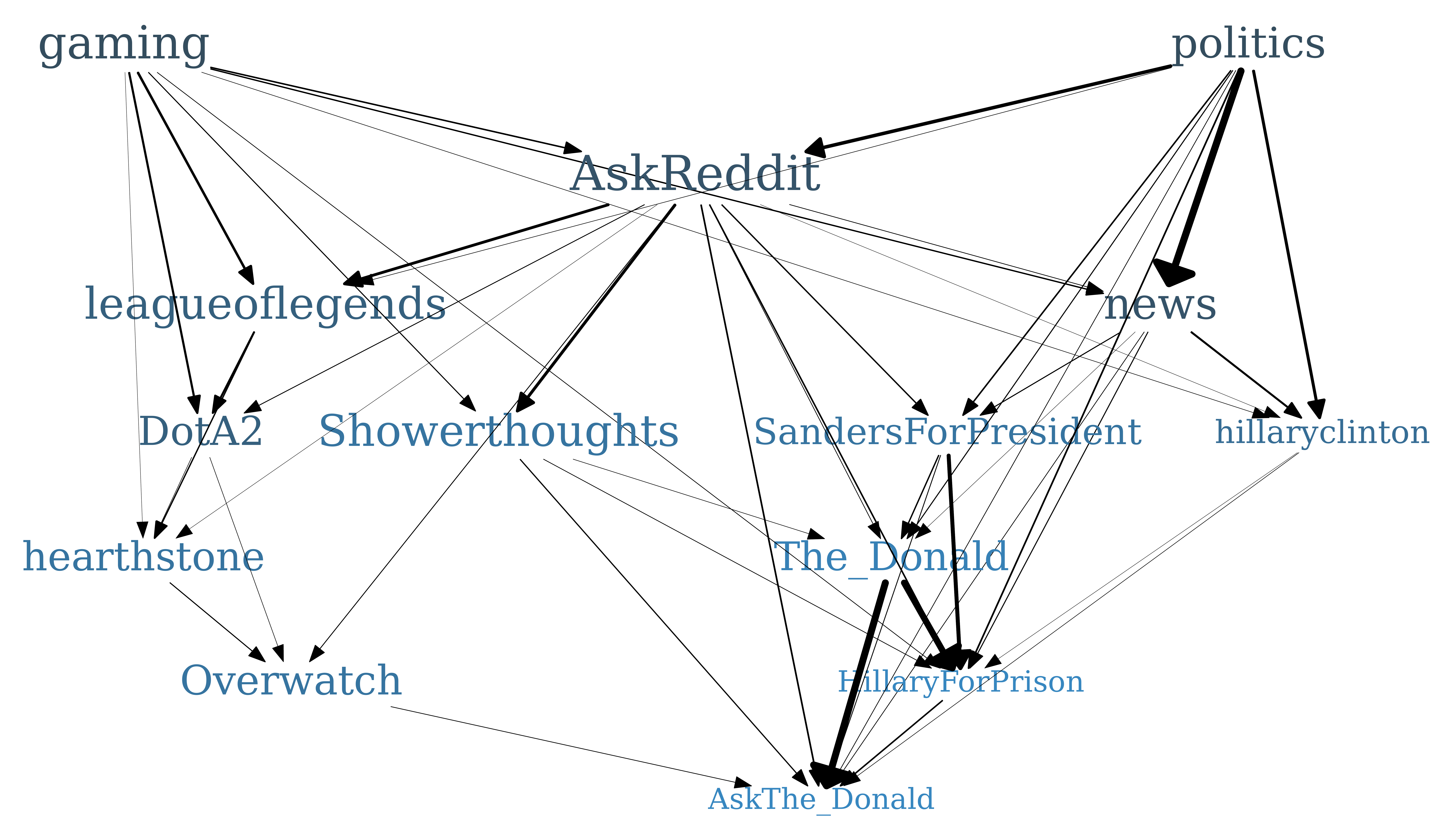}
        \caption{A subgraph of the genealogy graph}
        \label{fig:intro_100}
    \end{subfigure}
    \caption{Genealogy graphs of example communities on Reddit based on the first 100 members.
    A directed edge indicates that there exist early members of the target node (``child'' community) that were members of the source node (``parent'' community); the thickness (weight) of an edge represents the fraction of such members.
    Node color represents the depth in the genealogy graph (darker color for older communities), while node size indicates {\em community size} measured by the number of members.
    \figref{fig:intro_p1} and \figref{fig:intro_p2} show the top 10 parents of \communityname{AskThe\_Donald} and \communityname{Overwatch} respectively sorted by edge weights, while \figref{fig:intro_100} presents the genealogy graph between a sample of communities starting from two of the first communities on Reddit (\communityname{politics} and \communityname{gaming}) to \communityname{AskThe\_Donald} and \communityname{Overwatch}.
    To make the genealogy graph readable, we only present edges with weight greater than 0.01 (more than one members from the parent community).
    }
    \label{fig:intro}
\end{figure*}

\begin{flushright}
``We all carry, inside us, people who came before us.''\\
\rule[2.5pt]{10pt}{0.5pt} Liam Callanan
\end{flushright}

The tendency of individuals to flock together and form groups has led to continually emerging communities both online and offline.
Websites that allow users to create communities at their own discretion (e.g., Facebook, Reddit, 4chan) provide a great opportunity to document this trend.
For example, \figref{fig:sub_count} shows a rapid increase in the number of communities ever since Reddit allowed users to self-organize into topic-based communities.
Such growing trend poses an intriguing puzzle: where do these new communities come from?

In this work, we address this puzzle by considering each community as an entity, identifying the parents of a community, and building a genealogy of communities.\footnote{Anecdotally, GenWeekly has a series of articles on ``genealogy of communities'' \cite{gencomm}.}
Although a battery of studies have leveraged online communities to investigate group formation and community growth \cite{Backstrom:2006:GFL:1150402.1150412,Kairam:2012:LDO:2124295.2124374,kossinets2006empirical,LibenNowell:ProceedingsOfTheNationalAcademyOfSciences:2008,Pavlopoulou:PorceedingsOfThe12ThInternationalWorkshopOn:2017}, they tend to focus on a {\em single} community and seldom pay attention to the context that contains numerous existing communities.
Our idea resonates with studies that explain the emergence of organizations by analyzing the synergy between a small number of close existing communities \cite{padgett2012emergence}.
For instance, \citet{fleming2007valley} demonstrate how academic institutions and industry labs contribute to the emergence of high-tech companies in Silicon Valley and Boston.
Our work aims to provide a computational framework for tracing the origin of a community among {\em all} existing communities.
Our main observation is that although every community starts from scratch (0 members), it does not emerge from a vacuum.
In particular, members of a new community carry their own history, i.e., previous community memberships.
The history of the early members allows us to understand where a community comes from and as a result, trace the origin of a community.

\para{Examples from Reddit.} In order to build a genealogy of communities, we need the 
activity history of community members, which makes Reddit an ideal testbed.
\figref{fig:intro_p1} and \figref{fig:intro_p2} show the top 10 parents of \communityname{AskThe\_Donald} and \communityname{Overwatch} respectively, based on the first 100 members in each community.
Not surprisingly, \communityname{AskThe\_Donald}, a community where people ask Trump supporters questions, comes from \communityname{The\_Donald}, a community for Trump supporters, and other politics related communities such as \communityname{HillaryForPrison} and \communityname{SandersForPresident}.
Similarly, \communityname{Overwatch}, a community for a Blizzard game, comes from gaming communities such as \communityname{wow}, \communityname{hearthstone}, and \communityname{Diablo}.
We also notice differences in edge weights between these two examples.
Not any single parent of \communityname{Overwatch} is as dominant as \communityname{The\_Donald} for \communityname{AskThe\_Donald}.

\figref{fig:intro_100} presents a broader view of the entire genealogy graph starting from the first Reddit communities, such as \communityname{politics} and \communityname{gaming}. 
We make several interesting observations.
First, politics related communities and gaming related ones are clearly divided into two camps.
Second, the connections are denser and weighted heavier for politics related communities, e.g., (\communityname{The\_Donald}, \communityname{HillaryForPrison}).
Third, despite being created more recently than \communityname{politics} and \communityname{gaming}, some communities such as \communityname{AskReddit} and \communityname{Showerthoughts} are among the largest communities on Reddit.

In this work, we provide the first large-scale characterization of such genealogy graphs.
We further demonstrate the connection between the origin of a community and its future growth.
We finally investigate the characteristics of early members to understand the basic element of our genealogy graphs.

\para{Organization and highlights.}
We start by giving an overview of related work on group formation
to provide further background for this work.
We then introduce our dataset from Reddit, which spans over a decade.

We propose a framework for building genealogy graphs based on the first $k$ members.
Using this framework, we first track how a community {\em emerges} by examining the genealogy graphs when the number of members ($k$) grows from 0 to 100.
We find that as $k$ increases, the number of parents increases and the average weights of parents declines, indicating that the new community is becoming less dependent on any existing community.
Second, we investigate how basic graph properties of the genealogy graph evolve as new communities emerge on Reddit.
Intriguingly, despite the rapid increase in the number of communities over time, genealogy graph properties quickly converge to a stable state: the number of parents based on the first 100 members converges to $\sim$\avgparentsize and the dominant parent tends to contribute 10\% of the first 100 members.

We further investigate how the origin of a community connects to its future growth.
We find that genealogy graph information is useful for predicting how quickly community size grows (a 8.7\% relative improvement in mean squared error).
Our results show that strong parent connections are important for future community growth.
This finding suggests that the emerging process of a community is analogous to complex contagion, e.g., the diffusion of political hashtags requires dense connections between early adopters \cite{Centola:AmericanJournalOfSociology:,Fink:SocialNetworkAnalysisAndMining:2015,Romero:2011:DMI:1963405.1963503}.

Finally, we formulate a prediction problem to better understand 
the characteristics of early members {\em at the individual level}.
A diverse portfolio across existing communities turns out to be the most important predictor for becoming an early member, whereas community feedback and language use do not matter.
Studies on early adopters of new products have found an overlap between early adopters, opinion leaders, and market mavens, who have information about many kinds of products and markets \cite{Baumgarten:JournalOfMarketingResearch:1975,Feick:JournalOfMarketing:1987,Rogers:10}.
Our finding suggests that 
early members of a new community tend to be market mavens instead of opinion leaders.

\section{Related Work}
\label{sec:related}

Group formation and evolution has long been a focus in social science research \cite{lewin1951field,coleman1990foundations}.
Here we discuss two most relevant strands of literature to provide background for this work.

\para{Group formation as a diffusion process.}
The process of group formation can be viewed as a diffusion process where joining a new group is analogous to adopting an innovation.
A battery of studies have investigated the diffusion process of a single community or innovation \cite{Aral:ProceedingsOfTheNationalAcademyOfSciences:2009,Backstrom:2006:GFL:1150402.1150412,Bakshy:2012:RSN:2187836.2187907,Centola:Science:2010,Centola:AmericanJournalOfSociology:,Goel:ManagementScience:2015,kossinets2006empirical,LibenNowell:ProceedingsOfTheNationalAcademyOfSciences:2008}.
The seminal work on group formation by \citet{Backstrom:2006:GFL:1150402.1150412} shows that the likelihood of a person to join a group is associated with the number of her friends in that group;
\citet{Centola:AmericanJournalOfSociology:} propose the idea of complex contagion and suggest that the diffusion of certain behavior (e.g., political opinion) requires repeated contact and dense connections between early adopters.

Our work offers a new perspective in the context of constantly emerging communities: we view each community as an entity and focus on the relations between a new community and existing communities.
In comparison, the basic unit of analysis from the diffusion perspective is the individual and the context in which a community emerges are existing networks between individuals.

\para{Individual identity vs. community identity.}
Another closely related line of work is social identity \cite{tajfel2010social}.
One central hypothesis is that an individual's self-perception derives from her group memberships.
Thanks to the availability of datasets from online communities, researchers have become increasingly interested in studying user engagement in {\em multiple} communities as well as inter-group relations \cite{Hamilton:ProceedingsOfIcwsm:2017,tan2015all,Vasilescu:2014:SQS:2531602.2531659,zhang2017community,Zhu:2014:SEN:2556288.2557348,Zhu:2014:IMO:2556288.2557213,hessel2016science,kumar2018conflict}.
For instance, \citet{tan2015all} construct the life trajectory of individuals using the communities to which individuals post on Reddit, and show that continual exploration is connected to users' lifespan.

Instead of understanding individual identity from her life trajectory, our work attempts to examine the community identity by tracing where a community comes from. 
Community identity and individual identity are intertwined and their relation resonates with the two ecologies proposed in \citet{Astley:AdministrativeScienceQuaterly:1985}, population ecology and community ecology.
We take the community ecology perspective and investigate 
the emergence of communities.

Finally, although our work focuses on the {\em emergence} of communities using explicit community structures, it is related to studies on online community building and loyalty \cite{Hamilton:ProceedingsOfIcwsm:2017,hirschman1970exit,Kim:2000:CBW:518514,kraut2012building}, and works on implicit community detection \cite{Clauset:PhysRevEStatNonlinSoftMatter:2004,Girvan:ProceedingsOfTheNationalAcademyOfSciences:2002,Yang:ProceedingsOfWsdm:2013}.

\section{Reddit Communities} 
\label{sec:data}

Our main dataset is drawn from Reddit, a popular website where users can submit, comment on, upvote, and downvote posts \cite{Singer:Www14:2014,tan2015all}.
We refer to the difference between the number of upvotes and downvotes as {\em feedback} for a post.
We use posts from the inception of Reddit until April 2017.\footnote{This dataset is from \url{https://files.pushshift.io/reddit/}, thanks to J. Baumgartner.
A small amount of data is missing due to scraping errors and other unknown reasons \cite{gaffney2018caveat}.
We checked the sensitivity of our results to missing posts with a dataset from J. Hessel; our results do not change after accounting for them.
Details at \url{https://chenhaot.com/papers/community-genealogy.html}.
}

\para{A brief history of communities on Reddit.}
Reddit started with a main discussion forum, \communityname{reddit.com}, and other subreddits such as \communityname{science}, \communityname{politics}, and \communityname{gaming}.
In 2008, Reddit released a feature that allows users to create their own subreddits.\footnote{\url{https://redditblog.com/2008/01/22/new-features/}.}
Each subreddit focuses on a particular topic and has its own rules and norms, and thus functions as a community.
Reddit now consists of tens of thousands of subreddits (henceforth {\em communities}), and the fact that we have access to the posting history of users enables our study.

This paper focuses on communities with more than 100 members so that we have a reasonable size of user base to trace the genealogy.
As shown in \figref{fig:sub_count}, the number of communities has been increasing rapidly since 2008.
There are over 30,000 communities with more than 100 members until January 2017.
We consider a user as a member of a community if she made a post to that community.\footnote{We only consider subreddits created until January 2017 so that even the newest subreddits have 3 months to accumulate members.
Following \citet{tan2015all}, we only consider posting behavior and do not include commenting behavior because posting is initiated by users themselves, while commenting depends on other confounding factors such as the ranking system of Reddit.}
We refer to the number of users in a community as {\em community size}.

Our goal in this paper is to reveal the emergence of communities through building genealogy graphs between communities and exploring such genealogy graphs.
The rapid increase in \figref{fig:sub_count} is not even hindered by policies that raised the bar for creating communities by introducing additional user criteria in 2015.\footnote{For more details, see \url{https://www.reddit.com/r/help/comments/2yob6r/creating_a_subreddit/}. This rule only limits who can create a subreddit and does not affect early members.}
Such an ever-growing set of communities further motivates our research
to understand the emergence of communities.

\begin{figure*}[t]
    \begin{subfigure}[t]{0.28\textwidth}
        \addFigure{\textwidth}{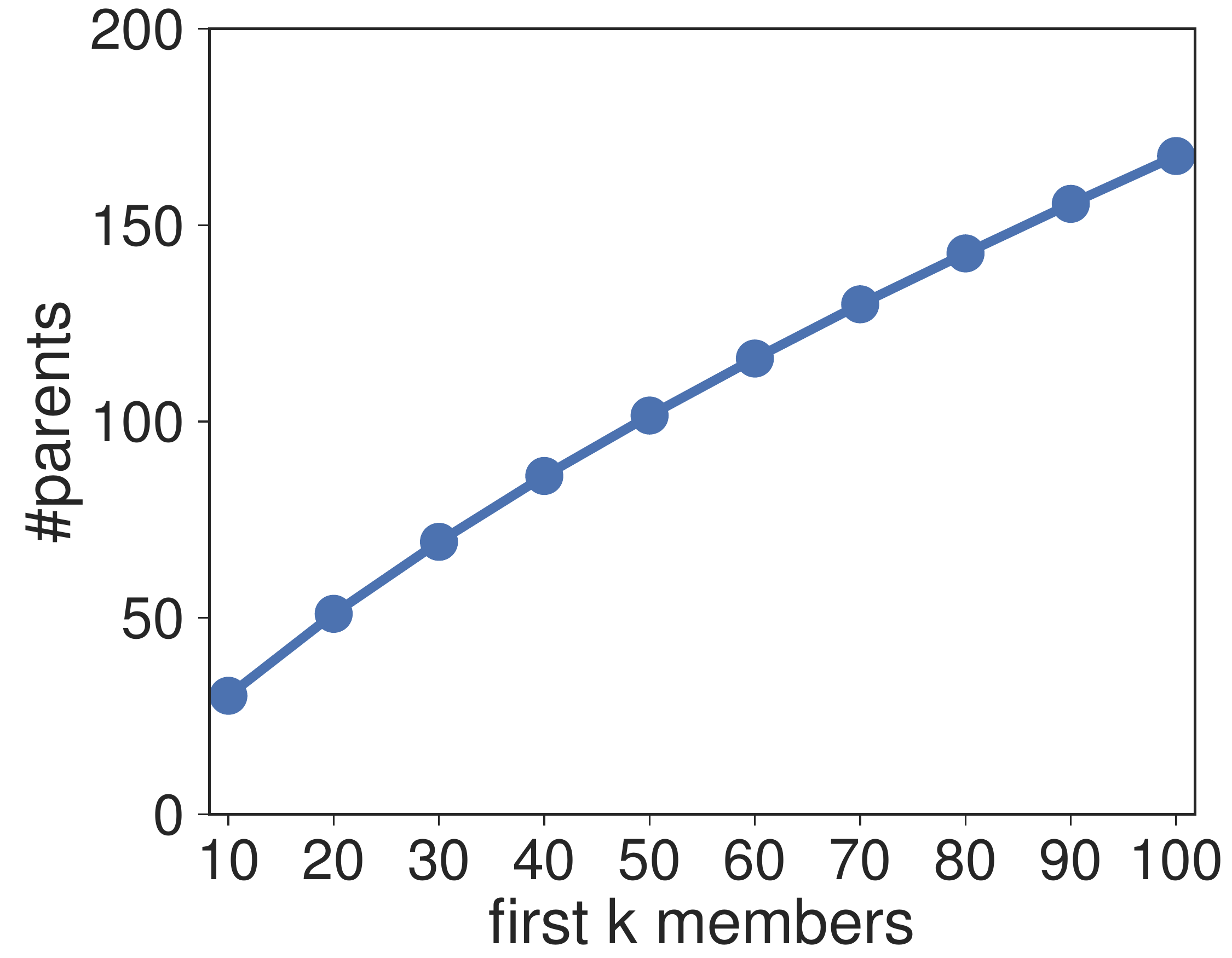}
        \caption{Number of parents}
        \label{fig:parent_trends}
    \end{subfigure}
    \hfill
    \begin{subfigure}[t]{0.28\textwidth}
        \addFigure{\textwidth}{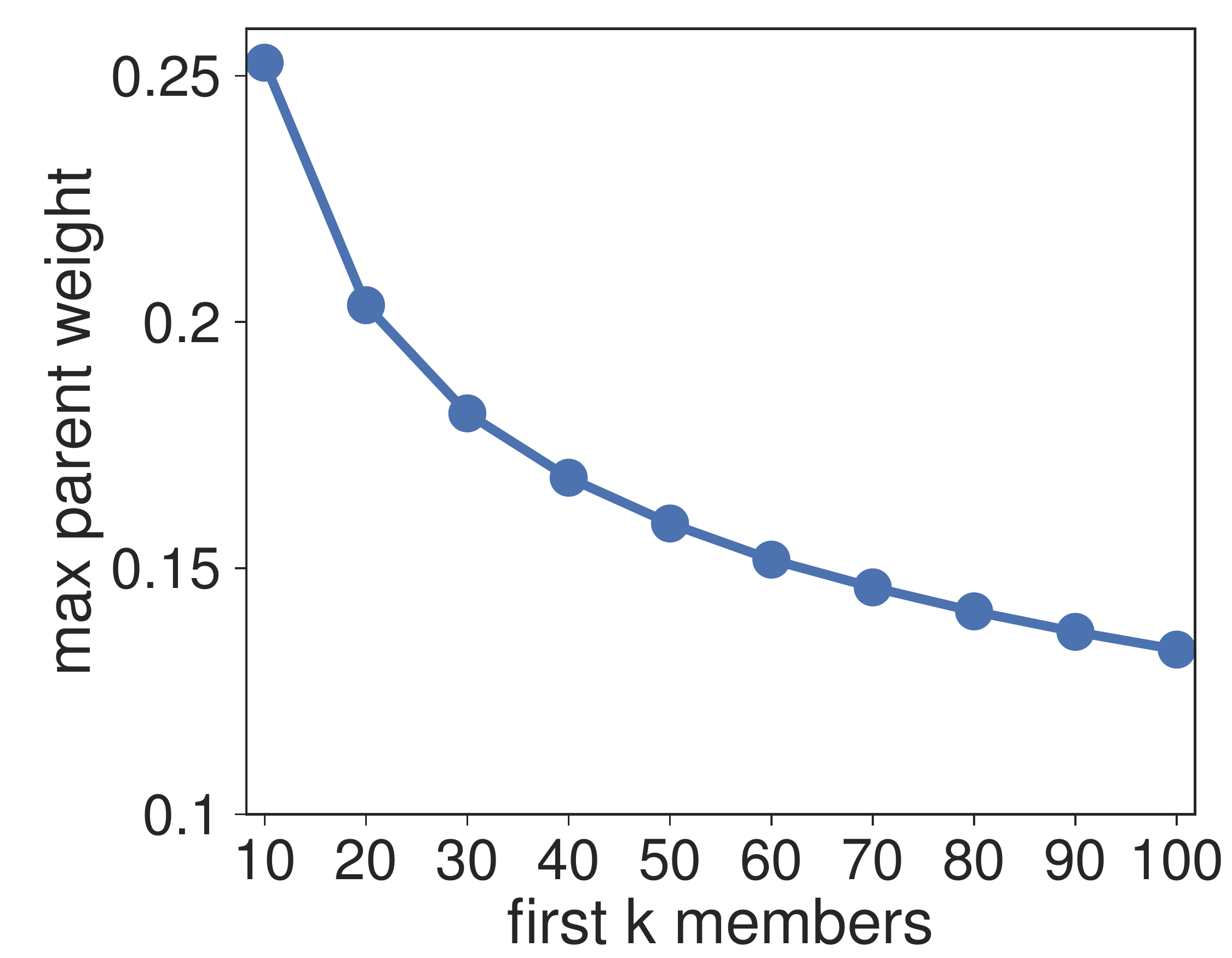}
        \caption{Max parent weight}
        \label{fig:max_weight_trends}
    \end{subfigure}
    \hfill
    \begin{subfigure}[t]{0.28\textwidth}
        \addFigure{\textwidth}{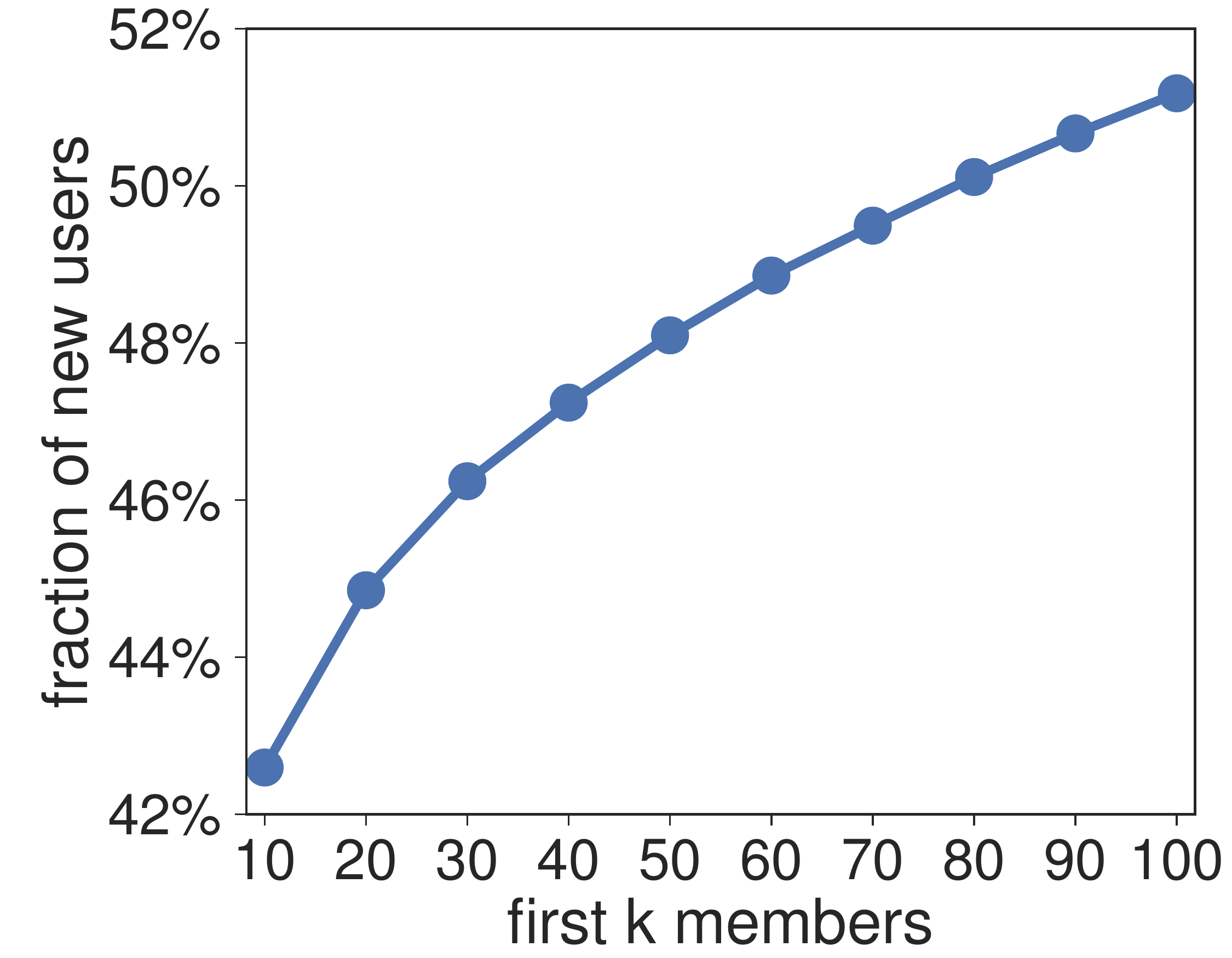}
        \caption{Fraction of new users}
        \label{fig:new_user_trends}
    \end{subfigure}
    \caption{
    x-axis represents the number of members and y-axis represents basic properties of genealogy graphs. Error bars (tiny) represent standard errors.
    As the number of members increases, the number of parents of a new community increases in the genealogy graph, while max parent weight declines. Meanwhile, the fraction of new users, who do not have any recent community membership, increases. We also group communities by creation year and by community size to address concerns regarding Simpson's Paradox and same trends hold in all conditions \cite{barbosa2016averaging} (this is also true for \figref{fig:relation_time}).
    }
    \label{fig:community_k}
\end{figure*}

\section{Characterizing Genealogy Graphs}
\label{sec:genealogy}

In this section, we introduce the definition of genealogy graphs and explore their basic properties.
We examine how genealogy graphs change as a community emerges, i.e., more users join the community.
We also investigate the evolution of genealogy graph properties over time as users create more communities and find an intriguing convergence.

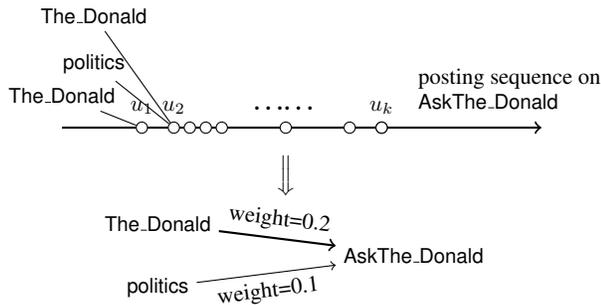
\begin{figure}[t]
\begin{center}
\begin{tikzpicture}[scale=0.85, every node/.style={scale=0.85}]
  \draw[thick,->] (-1, 0) -- (6.5, 0);
  \node[circle,draw=black, fill=white, inner sep=0pt,minimum size=5pt, label={$u_1$}] (u1) at (0.25, 0) {};
  \node[circle,draw=black, fill=white, inner sep=0pt,minimum size=5pt, label={$u_2$}] (u2) at (0.75, 0) {};
  \node[circle,draw=black, fill=white, inner sep=0pt,minimum size=5pt] (u3) at (4, 0) {};
  \node[circle,draw=black, fill=white, inner sep=0pt,minimum size=5pt] (u4) at (1, 0) {};
  \node[circle,draw=black, fill=white, inner sep=0pt,minimum size=5pt] (u5) at (1.25, 0) {};
  \node[circle,draw=black, fill=white, inner sep=0pt,minimum size=5pt] (u6) at (1.5, 0) {};
  \node[circle,draw=black, fill=white, inner sep=0pt,minimum size=5pt, label={$\boldsymbol{\cdots\cdots}$}] (u6) at (2.5, 0) {};
  \node[circle,draw=black, fill=white, inner sep=0pt,minimum size=5pt] (u6) at (3.5, 0) {};
  \node[circle,draw=black, fill=white, inner sep=0pt,minimum size=5pt, label={$u_k$}] (uk) at (4, 0) {};
  \node[label={[align=left]posting sequence on \\\communityname{AskThe\_Donald}}] (time) at (6, 0.05) {};

  \node (s1) at (-1, 0.5) {\communityname{The\_Donald}};
  \node (s2) at (-0.5, 1.75) {\communityname{The\_Donald}};
  \node (s3) at (-0.5, 1) {\communityname{politics}};

  \draw (u1) -- (s1);
  \draw (u2) -- (s2);
  \draw (u2) -- (s3);

  \node (arrow) at (2.5, -0.75) {$\Big\Downarrow$};
  \node (askdonald) at (4.5, -2) {\communityname{AskThe\_Donald}};
  \node (donald) at (0.5, -1.5) {\communityname{The\_Donald}};
  \node (politics) at (0.5, -2.5) {\communityname{politics}};
  \draw[thick, ->] (donald) -- (askdonald) node [midway, above, sloped] {weight=0.2};
  \draw[->, label={0.1}] (politics) -- (askdonald) node [midway, below, sloped] {weight=0.1};
\end{tikzpicture}
\end{center}
\caption{Illustration of the genealogy building algorithm for \communityname{AskThe\_Donald}. The timeline shows the early members of \communityname{AskThe\_Donald} and their posting sequence.
Using the recent posting history of $u_1$ and $u_2$, we can compute the weights for the genealogy graph at $k=10$, assuming $u_3, \dots, u_{10}$ did not make any posts in the previous month.}
\label{fig:demo}
\end{figure}

\subsection{Building Genealogy Graphs}
\label{sec:genealogy_definition}

The key to building a genealogy of communities is to find a set of ``parent'' communities for each new community.
We trace the parents of an emerging community by examining where its early members were active right before joining the new community.
In other words, we construct a community's identity based on its early members' {\em recent} community memberships.
Specifically, we define parents of a new community $j$ based on the posting history of its first $k$ members in the month before they posted to community $j$.
\figref{fig:demo} presents an example for \communityname{AskThe\_Donald} when $k=10$.
We define the weight on edge $ij$ as the fraction of early members in $j$ that were active in $i$:
$$w_{ij} = \frac{|\{u_r| 1 \leq r \leq k, u_r \text{ posted in i recently}\}|}{|k|},$$
\noindent where $i$ is the {\em parent} community, $j$ is the {\em child} community, and $u_r$ is the $r^{\text{th}}$ member in community $j$.
We identify the parents of all communities that were created after 2008, when users started to self-organize into communities.
We approximate the creation time of a community by when the first post was made to a community.
We only keep edge $ij$ if community $i$ was created before community $j$.
To measure how dominant the top parent is (e.g., \communityname{The\_Donald} for \communityname{AskThe\_Donald} in \figref{fig:intro_p1}),
we define {\em max parent weight} as $\max_i w_{ij}$ for child community $j$.

The number of early members ($k$) is an important parameter in this definition.
We use small $k$ values to focus on the emergence of a community when the total number of members is small ($\leq 100$).
For the same reason, we use an absolute count of $k$ instead of a relative percentage.
As $k$ increases, a community's identity takes shape. 
We track the emergence of a community by varying $k$ from 10 to 100 and observe how genealogy graphs change.

{\em Our approach is the first attempt to build genealogy graphs between communities.
We will discuss future research directions in the concluding discussion.
The extracted genealogical edges necessarily depend on our definition through previous community memberships of early members, so parent-child relations in our genealogy graphs indicate where the ``child'' comes from.
It follows that ``parent'' and ``child'' do not have to relate semantically, e.g., \communityname{Overwatch} being a ``parent'' of \communityname{AskThe\_Donald} in \figref{fig:intro_p1}.}

\begin{figure*}
    \begin{subfigure}[t]{0.3\textwidth}
        \addFigure{\textwidth}{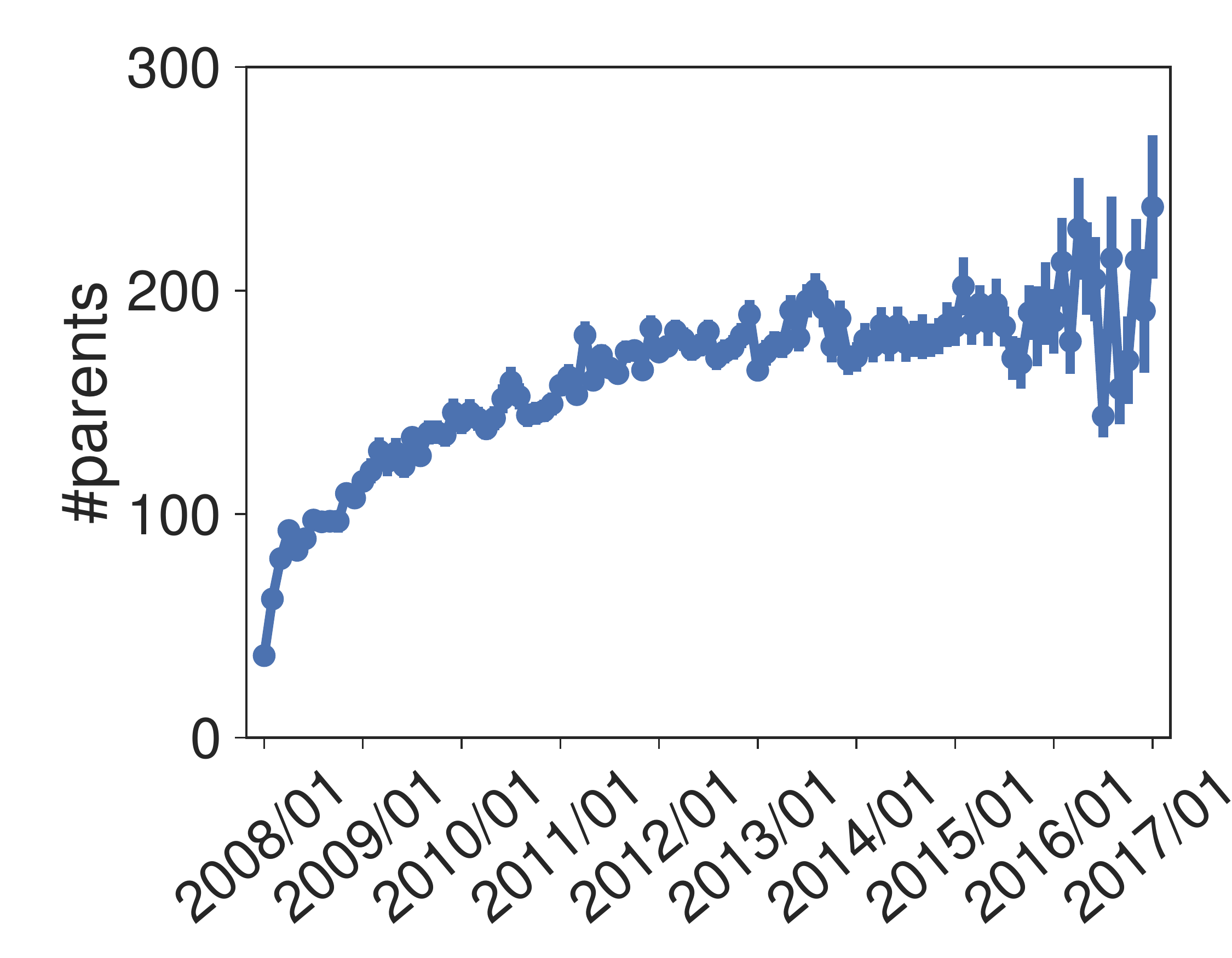}
        \caption{Number of parents}
        \label{fig:num_parents_time}
    \end{subfigure}
    \hfill
    \begin{subfigure}[t]{0.3\textwidth}
        \addFigure{\textwidth}{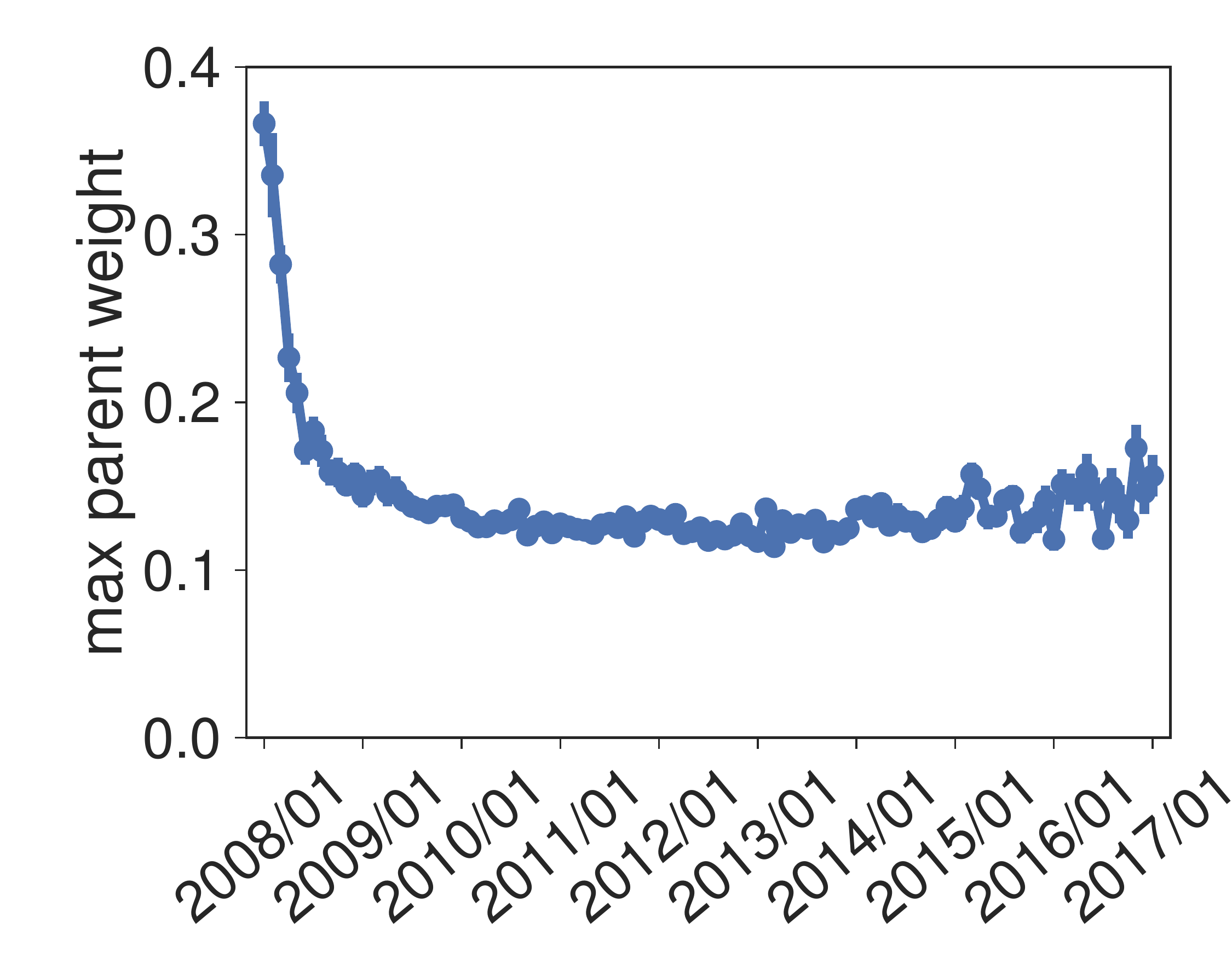}
        \caption{Max parent weight}
        \label{fig:max_weight_time}
    \end{subfigure}
    \hfill
    \begin{subfigure}[t]{0.3\textwidth}
        \addFigure{\textwidth}{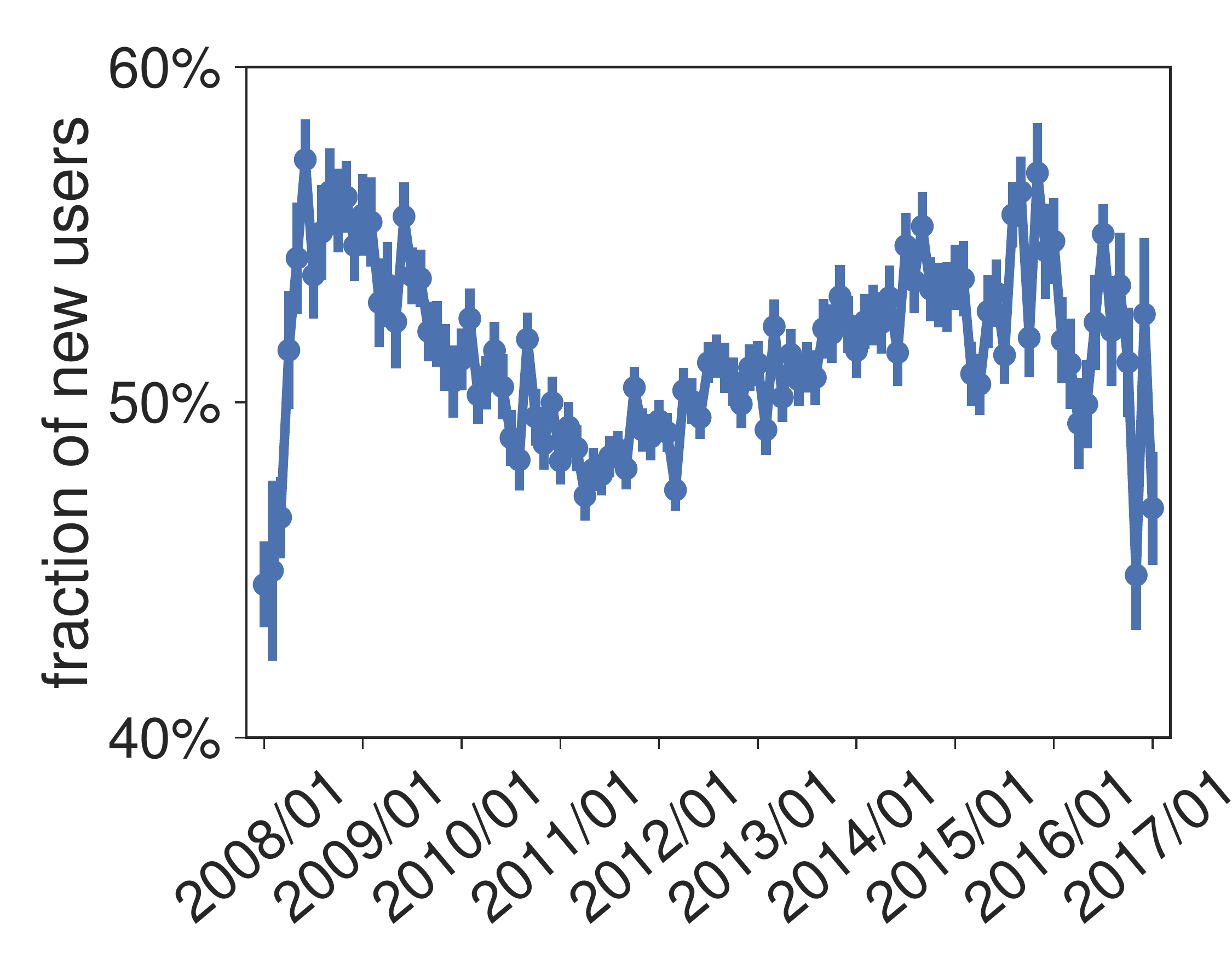}
        \caption{Fraction of new users}
        \label{fig:new_user_time}
    \end{subfigure}
    \caption{Evolution of genealogy graph properties over time. Each point represents the average corresponding properties of communities created in that month, and error bars represent standard errors. Both the number of parents and max parent weight converge rather quickly despite the rapid increase in the number of communities.
    }
    \label{fig:relation_time}
\end{figure*}

\subsection{From 0 Members to 100 Members}

As the saying goes, ``Rome wasn't built in one day'', every community starts from 0 members no matter how many members it eventually has.
We study the emerging process by tracking the genealogy graph from $k=10$ to 100.

We study how basic properties of genealogy graphs between communities change in the emerging process of a community (\figref{fig:community_k}).
As the community matures and gains more members, additional members bring more past history and thus more parent communities in the genealogy graph.
Meanwhile, the influence of parents decreases, indicated by the declining trend of max parent weight.
The declining influence of parents is inherent to the construction of the child community's identity.

Another important role that a new community plays is to attract new users who do not have any previous community membership.\footnote{This includes both new users on Reddit and ``old'' Reddit users who did not post in the one month before posting in the new community, which can be considered reactivated ``new'' users.}
We observe an increasing fraction of new users as a community emerges:
earlier members (small $k$) tend to come with past history, while a larger fraction of later members (larger $k$) come afresh.

\subsection{Dynamics over Time}

In addition to changes during the emergence of a community, we investigate how genealogy graph properties evolve over time as the Reddit website as a whole grows.
Despite the rapid increase in the number of communities (\figref{fig:sub_count}), graph properties converge to a stable state rather quickly.
We focus our discussion on $k=100$, but the observations are robust across choices of $k$.

\para{The number of parents grows and converges to around \avgparentsize (\figref{fig:num_parents_time}).}
As there are more communities over time and users continually explore new communities \cite{tan2015all}, the number of parents increases because users carry more previous community memberships.
However, the growth seems to have stopped since 2012.
For communities created after 2012, the first 100 members tend to be previously active in around \avgparentsize existing communities.

\para{Max parent weight declines and converges to 0.1 (\figref{fig:max_weight_time}).}
As the number of communities increases on Reddit, early members of a new community tend to come from more diverse backgrounds.
Therefore, max parent weight declines from $\sim$0.4 to 0.1.
However, this decline happens rather quickly.
Starting from early 2009, max parent weight stabilizes at 0.1, suggesting that on average 10\% of the first 100 members of a new community come from the same parent community. 
This observation holds despite that the number of existing communities grows from 100s to 10,000s since 2009.

\para{The fraction of new users fluctuates over time (\figref{fig:new_user_time}).}
Unlike the number of parents and max parent weight, the fraction of new users presents a puzzling shape.
The fraction of new users increased in early 2008 as Reddit just started to allow user-created subreddits.
However, it went through a declining period until around 2011 and then started to grow again.
One possible explanation for the increase in 2011 is the shutdown of the original main-reddit, \communityname{reddit.com}, which may reactivate some users to explore new communities, but this does not suffice to explain the growth in the fraction of new users from 2011 to 2014.

\section{Predicting Community Growth}
\label{sec:success}

Having established the temporal dynamics of genealogy graphs between communities, we now investigate how the origin of a community relates to its future growth.
We formulate a prediction framework and study a community's future growth after its number of members reaches 100.
Our results consistently show that strong parent connections are associated with future community growth.

\subsection{Problem Setup}

In order to study how the origin of a community relates to its future growth, we develop two prediction tasks towards future community growth.
First, inspired by \citet{Cheng+etal:2014} on predicting cascades, we use the median community size in our sample to characterize a community's future growth and predict if a community's future size is going to exceed the median community size (\mediancommunitysize in our dataset).\footnote{Different from the observation in \citet{Cheng+etal:2014}, although community size follows a heavy-tailed distribution, the hypothesis that it follows a power-law distribution is rejected when testing the goodness-of-fit \citep{clauset2009power}. We thus use the empirical median instead of an estimated value from the best power-law fit.}
The advantage of this prediction setup is twofold: 1) it controls for the current size of a community and depends entirely on future community size; 2) it naturally leads to a balanced classification task.
The majority baseline gives an accuracy of around 50\%.
We call this task {\em growth classification} for short.

The second task examines the rate of growth.
We estimate how long it takes to reach the median for the subset of communities that exceed \mediancommunitysize members.
We use $\log(t_{\mediancommunitysize} - t_{100})$ as the target variable and formulate a regression problem, where $t_k$ indicates the first posting time of the $k^{\textrm{th}}$ member.
This regression task focuses on about half of the communities in the previous classification task.
We refer to this task as {\em rate regression}.

For both tasks, in addition to using all the information from a community's first 100 members, we constrain ourselves to using only the information from the first $k=10, 20, \dots, 90$ members so that we can evaluate the marginal gain from additional members' information.

\subsection{Characterizing the Origin of a Community}

Inspired by previous work on community growth \cite{Ducheneaut:ProceedingsOfChi:2007,Kairam:2012:LDO:2124295.2124374,Zhu:2014:SEN:2556288.2557348,Zhu:2014:IMO:2556288.2557213}, we consider the following hypotheses and features to characterize the origin of a community.
Since both a large future size and a fast growth rate indicate community growth,
a feature that is positively correlated with whether the future community size exceeds the median should be negatively correlated with growth rate.
We first report significance testing results for single features when $k=100$ (\tableref{tb:community_prediction}) and then examine the effectiveness of these features in prediction for $k = 10, 20, \dots, 100$.

\para{Temporal features.} We employ the community creation time and the average time gap between posts as features.
We use the community creation time to account for the fact that the Reddit website has been growing.
It is less likely for newer communities to exceed the median size on average (partly because of a shorter history), however, given that the community size exceeds the median size, the growth takes less time for newer communities.
We expect smaller average time gap to be positively correlated with future community growth and this is confirmed in \tableref{tb:community_prediction}.
These two features constitute a strong baseline in previous studies \cite{Cheng+etal:2014,Kairam:2012:LDO:2124295.2124374}.

\para{Basic parent properties.}
The next two feature sets derive from the genealogy graph.
We first capture basic parent properties by the number of parents and edge weights.
We also include the interaction of these two variables, i.e., the number of parents with weights at least 0.05 or 0.1.
A large number of parents and heavy parent weights indicate strong parent connections, which may make a community more likely to grow.
However, diffusion studies suggest that large-scale diffusion can happen with very shallow depth and involve many individuals' independent adoption \cite{Bakshy:2011:EIQ:1935826.1935845,Goel:ManagementScience:2015,Romero+Tan+Ugander:13}.
Our results show that max parent weight is associated with future community growth, both in growth classification and rate regression.
The number of parents with substantial weights ($\geq 0.1$) is positively correlated in growth classification,
whereas the number of parents is {\em negatively} correlated.
These observations suggest that {\em strong parent connections help community growth but simply having many parents can sometimes hurt}.

\begin{figure*}[t]
    \centering
    \begin{subfigure}[t]{0.3\textwidth}
        \addFigure{\textwidth}{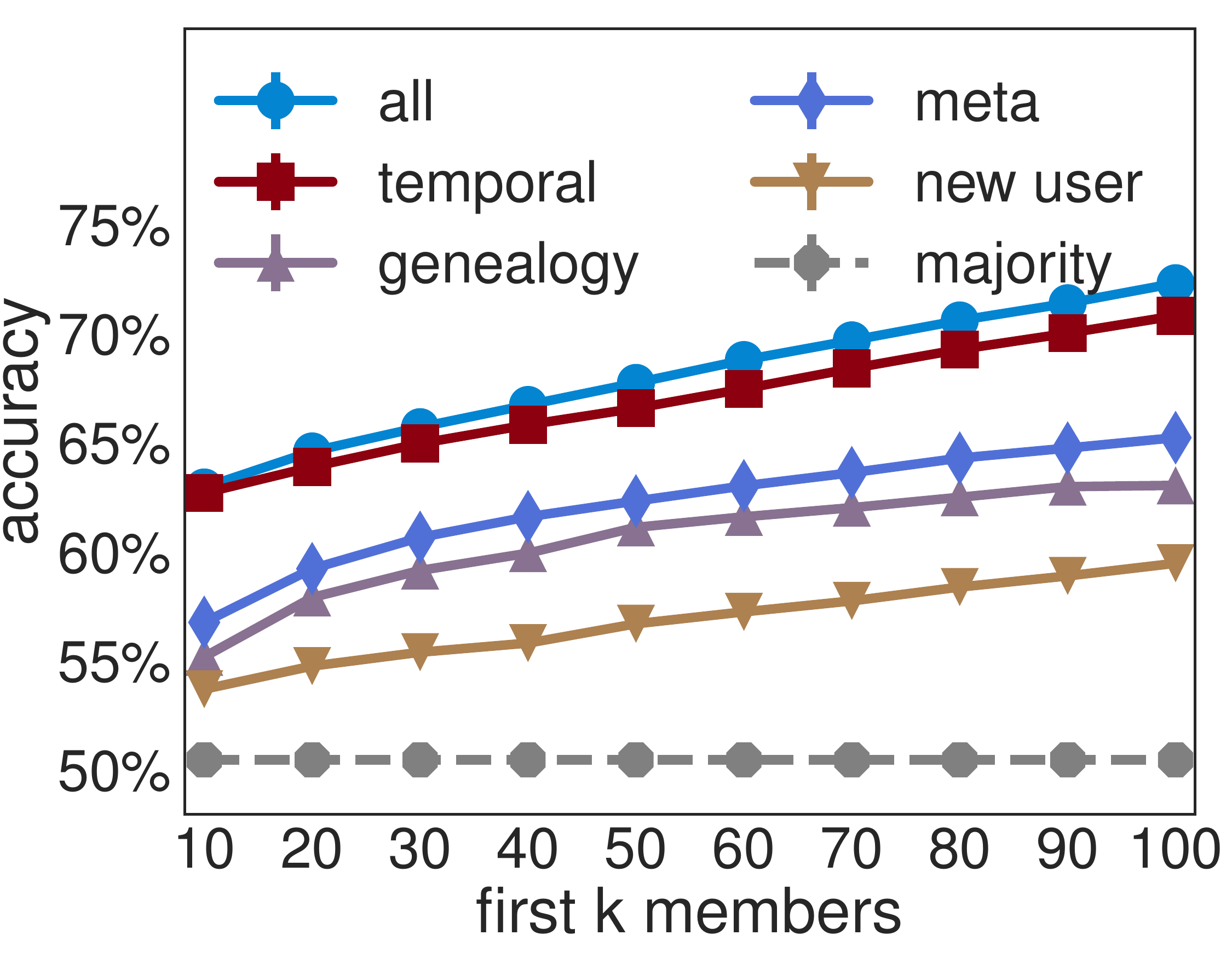}
        \caption{Growth classification performance.}
        \label{fig:cp_prediction}
    \end{subfigure}
    \hfill
    \begin{subfigure}[t]{0.3\textwidth}
        \addFigure{\textwidth}{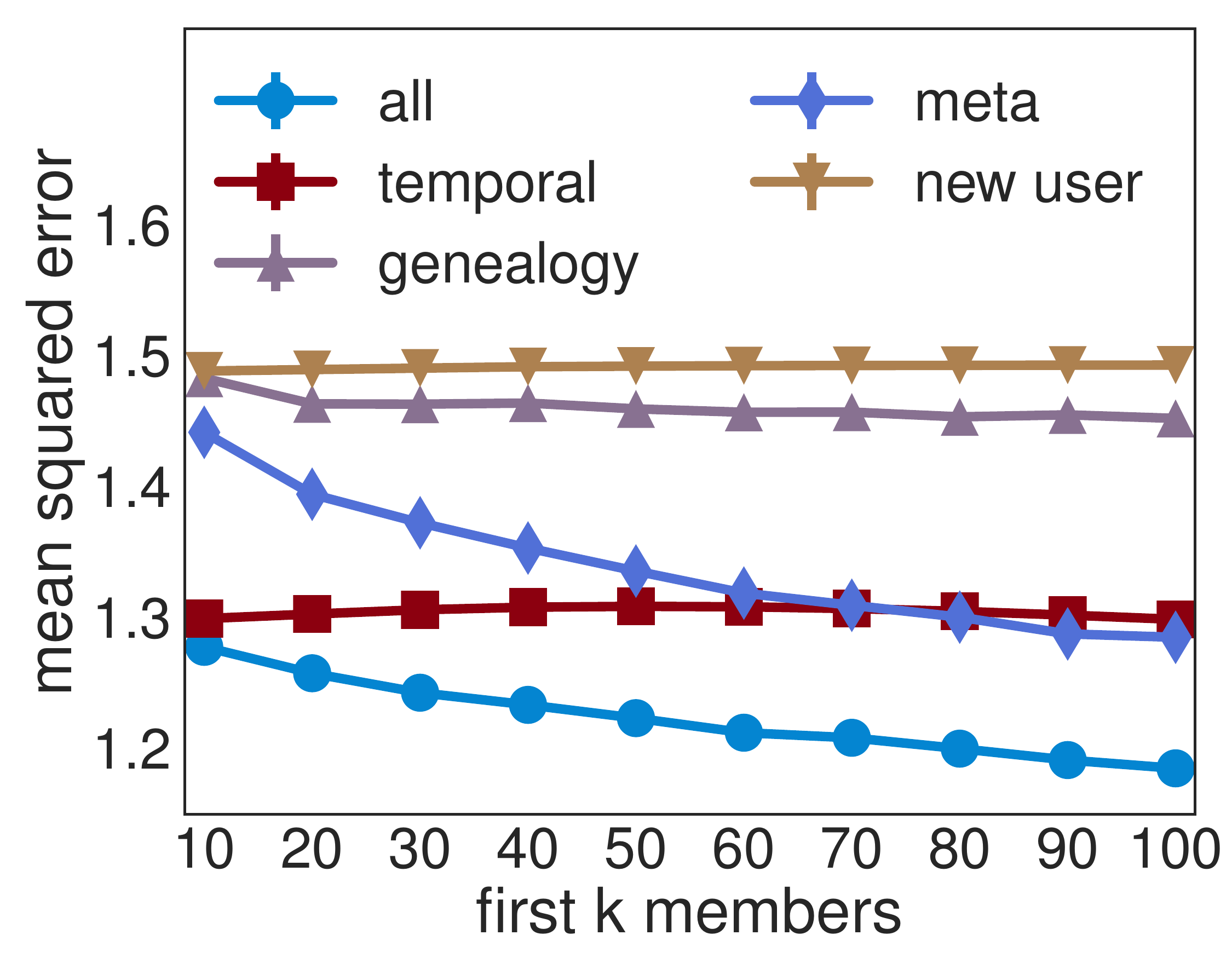}
        \caption{Rate regression performance.}
        \label{fig:cp_regression}
    \end{subfigure}
    \hfill
    \begin{subfigure}[t]{0.3\textwidth}
        \addFigure{\textwidth}{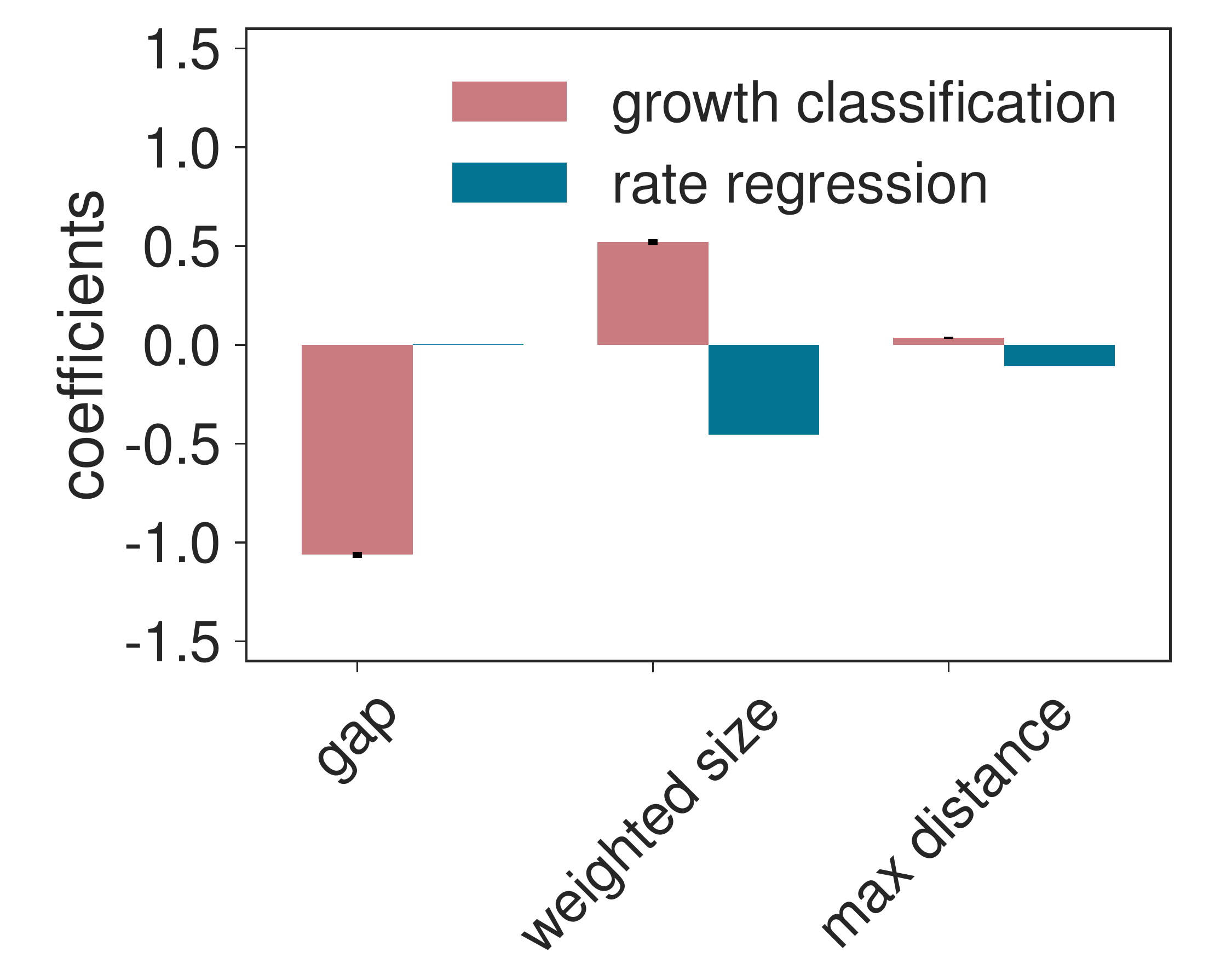}
        \caption{Coefficient comparisons.}
        \label{fig:cp_coefficients}
    \end{subfigure}
    \caption{
    \figref{fig:cp_prediction} shows the prediction accuracy for different feature sets as the number of members grows (the higher the better),
    while \figref{fig:cp_regression} presents the mean squared error (MSE), and smaller MSE indicates better performance.
    \figref{fig:cp_coefficients} compares feature coefficients of these two tasks when $k=100$ (features are standardized).
    Error bars represent standard errors.
    \label{fig:community_prediction}}
\end{figure*}

\para{Meta information of parents.} 
We also dive deeper into the origin of a community and measure meta information of parents.
We measure the meta information from two perspectives: size and language.
We compute the average, min, max, and standard deviation of parent sizes using $\log(\text{number of members})$ in the one month before the new community is created.
We expect larger parent sizes to associate with child community growth.
Our results show that not all parents matter the same way.
Average parent size is not significantly correlated in growth classification, but weighted average behaves as expected (larger weighted average parent size indicates larger likelihood to exceed the median community size and faster growth rate).

We measure pairwise distance between language models of the parent communities in the one month before the child community is created, using titles and texts in the posts.\footnote{We compute pairwise distance for only the top 20 weighted parents for efficiency reasons.
We also require that there are at least 100 unique members in that month to make sure that the language model is not too sparse.}
We compute the average, max, and standard deviation of pairwise distance and consider large language distance as a sign of diversity.
We expect a diverse set of parents to be associated with community growth because \citet{Uzzi:Science:2013} show that atypical combination is related to scientific impact.
However, this notion of diversity only seems effective for growth rate using max language distance.
Large average language distance hurts the likelihood of exceeding the median community size and has no effect on growth rate.
These results suggest that 
a closely related set of parents and maybe occasionally a small number of very different parents are associated with future community growth.

\begin{table}[t]
\small
\begin{tabular}{l@{}r@{\hskip 4pt}r}
\toprule
feature & growth classification & rate regression \\
\midrule
\multicolumn{3}{c}{temporal information} \\
community creation time & $\downarrow\downarrow\downarrow\downarrow$ & $\downarrow\downarrow\downarrow\downarrow$\\
average time gap & $\downarrow\downarrow\downarrow\downarrow$ & $\uparrow\uparrow\uparrow\uparrow$\\
\midrule
\multicolumn{3}{c}{basic parent properties} \\
\#parents & $\downarrow\downarrow\downarrow\downarrow$ & ------ \\
\#parents with weight $\geq 0.1$ & $\uparrow\uparrow\uparrow\uparrow$ & $\uparrow\uparrow$\\
max parent weight & $\uparrow\uparrow\uparrow\uparrow$ & $\downarrow\downarrow\downarrow\downarrow$\\
\midrule
\multicolumn{3}{c}{meta information of parents} \\
average parent size & ------  & $\downarrow\downarrow\downarrow\downarrow$\\
weighted average parent size & $\uparrow\uparrow\uparrow\uparrow$ & $\downarrow\downarrow\downarrow\downarrow$\\
average language distance & $\downarrow\downarrow\downarrow\downarrow$ & ------ \\
max language distance & $\downarrow\downarrow\downarrow\downarrow$ & $\downarrow\downarrow\downarrow\downarrow$\\
\midrule
\multicolumn{3}{c}{new user} \\
fraction of new users & $\downarrow\downarrow\downarrow\downarrow$ & ------ \\
\bottomrule
\end{tabular}

\caption{Testing results for community growth. We show a subset of features for space reasons. $\uparrow$ in growth classification and $\downarrow$ in rate regression are positive signals for community growth.
The number of arrows indicate $p$-values:
$\uparrow\uparrow\uparrow\uparrow: p < 0.0001$,
$\uparrow\uparrow\uparrow: p < 0.001$, $\uparrow\uparrow: p
< 0.01$, $\uparrow: p < 0.05$, the same for downward arrows; $p$ refers to the $p$-value after the Bonferroni correction. $t$-test is used for growth classification, while Pearson correlation is used for rate regression.
}
\label{tb:community_prediction}
\end{table}

\para{Fraction of new users.} 
The fraction of new users is another way to capture the broadcast style diffusion \cite{Bakshy:2011:EIQ:1935826.1935845,Goel:ManagementScience:2015,Romero+Tan+Ugander:13}.
We expect that the fraction of new users is positively correlated with future community growth.
However, \tableref{tb:community_prediction} shows the opposite.
This observation further suggests that it is important for a new community to have strong roots in existing communities.

\para{Summary.} Overall, our feature testing results consistently suggest that having strong parent connections in the genealogy graph is positively correlated with future community growth.
``Strong'' parent connections can be reflected by a heavy edge weight, large weighted average size, small average language distance, and a small fraction of new users.
The emerging process of a community is thus analogous to complex contagion \cite{Centola:AmericanJournalOfSociology:,Fink:SocialNetworkAnalysisAndMining:2015,Romero:2011:DMI:1963405.1963503} and requires dense connections between parents.

\subsection{Prediction Performance}
\label{sec:community_prediction_perf}

We evaluate the predictive power of all features in both growth classification and rate regression.
For each prediction experiment, we randomly split the data for training (70\%), validation (10\%), and testing (20\%).
We standardize each feature based on training data and use logistic regression with $\ell_2$-regularization.
We grid search 
the best $\ell_2$ coefficient based on performance on the validation set over $\bigl\{2^x\bigr\}$, where $x$ ranges from $-8$ to $1$.
We repeat this process 30 times and use Wilcoxon signed rank test to compare the performance of different feature sets \cite{Wilcoxon:BiometricsBulletin:1945}.

\para{Performance in growth classification (\figref{fig:cp_prediction}).}
Our classifier using all features outperforms the majority baseline and the performance improves quite significantly as $k$ grows ($\sim$10\% absolute gain in accuracy by having the first 100 members vs. the first 10 members).
The strongest single feature set is temporal information, which echoes the results in predicting cascades \cite{Cheng+etal:2014}.
The performance gap between using all features and using temporal information increases as $k$ grows, and the difference at $k=100$ is statistically significant (72.3\% vs. 70.8\%, $p < 0.0001$).
Meta information of parents is the second strongest feature set, whereas the fraction of new users performs the worst.
This observation further confirms the importance of a community's origin and suggests that the emerging process of a community depends less on broadcast-style diffusion in which many people join the group independently.

\para{Performance in rate regression (\figref{fig:cp_regression}).} 
Similar to the classification task, the performance improves as we get more information regarding early members (as $k$ increases).
However, different trends exist when comparing feature sets.
Although temporal information remains a useful single feature set, its performance does not improve as $k$ grows.
There is a 
8.7\% relative gain in mean squared error for $k=100$ by introducing features from the origin of a community (1.30 vs. 1.19).
Our results suggest that the origin of a community is more effective in rate regression than in growth classification.
One possible explanation is that the binary classification task is simple enough that temporal gap can already capture most of the signals.

\para{Feature coefficients comparisons (\figref{fig:cp_coefficients}).}
In order to further understand the difference between growth classification and rate regression, we compare the coefficients in the linear classifier/regressor.
The importance of average time gap drops significantly in the regression task (the coefficient's absolute value drops from 1 to almost 0), but similar drop does not happen for features based on the origin of a community such as weighted average parent size and max language distance (as expected, their feature coefficients have different signs in these two tasks).

\section{Who Becomes an Early Member?}
\label{sec:founder}

We finally turn to the individual level and examine the characteristics of early members, because they are central to the emergence of a community.
We formulate a prediction problem to differentiate
early members of a new community from another user that was similarly active in a parent community.

Although early members have not been systematically studied as it is rare to have such activity sequence data that documents a community's emerging process, many studies have looked at early adopters of new innovations and found an overlap between early adopters, opinion leaders, and market mavens \cite{Abratt:JournalOfBusinessAndPsychology:,Baumgarten:JournalOfMarketingResearch:1975,Feick:JournalOfMarketing:1987,Rogers:10,Ruvio:PsychologyMarketing:2007}.
Market mavens have information about many kinds of products and markets, while opinion leaders provide information and leadership in specific products.
Our results show that early members tend to be market mavens instead of opinion leaders.

\subsection{Problem Setup}

As active users may join a new community simply because of their high activity levels, we focus on understanding how factors other than activity levels relate to a user's decision to become an early member at a new community.
To do that, 
given a (parent, child) tuple in the genealogy graph and an early member of the child community (``positive''),
we find a matching user with similar activity level\footnote{The nearest neighbor in \#posts in the parent community in the month before the positive user joined the child community. We filter matches with distance greater than 5 \cite{stuart2008best}.} in the parent community but did not become an early member of the child community (``negative'').
The matching process leads to a balanced classification task, where the majority baseline gives an accuracy of 50\%.
We vary the number of early members ($k$) from 10 to 100 to examine 
how early members differ as a new community emerges.
We randomly sample 10K (parent, child) tuples, from which we collect 27.6K positive and negative users in total when $k=100$.

\begin{table}[t]
\begin{tabular}{p{0.27\textwidth}rr}
\toprule
feature & \multicolumn{2}{c}{significance} \\
\midrule
& parent & global \\
\#posts & ------ & $\uparrow\uparrow\uparrow\uparrow$ \\
average time gap & $\uparrow\uparrow\uparrow\uparrow$ & $\downarrow\downarrow\downarrow\downarrow$ \\
feedback &  ------ & ------ \\
language distance & ------  & $\downarrow\downarrow\downarrow\downarrow$ \\
std language distance & ------   & $\uparrow\uparrow\uparrow\uparrow$\\
\midrule
& interplay &\\
fraction of posts in parent & $\downarrow\downarrow\downarrow\downarrow$ & N/A \\
community entropy & $\uparrow\uparrow\uparrow\uparrow$ & N/A \\
\bottomrule
\end{tabular}

\caption{Testing results for predicting early members.
Upward arrows indicate positive correlation with becoming an early member, while downward arrows suggest the other way around
          ($\uparrow\uparrow\uparrow\uparrow: p < 0.0001$,
          $\uparrow\uparrow\uparrow: p < 0.001$, $\uparrow\uparrow: p
          < 0.01$, $\uparrow: p < 0.05$, the same for downward arrows; $p$ refers to the $p$-value after the Bonferroni correction).
          Since behavioral features are used for the parent community alone and the entire Reddit website, they are respectively reported in the ``parent'' and ``global'' column.
          }
\label{tb:founder_prediction}
\end{table}

\subsection{Individual Characteristics}

We design features based on an individual's behavioral data both only in the parent community and in the entire Reddit website.
Our features are derived from the following behavioral information: 

\begin{itemize}[topsep=2pt,leftmargin=*,itemsep=0pt]
    \item {\em Number of posts.} 
    Although we try to control for \#posts in the parent community, we expect more posts in Reddit to be positively correlated with becoming an early member.

    \item {\em Average time gap between posts.} More frequent posts indicate being more active and more likely to become an early member. 
    \item {\em Community feedback.} 
    Following \citet{tan2015all}, we measure community feedback by comparing a post's feedback with the median feedback in that month to control for the differences across communities.
    Positive community feedback can be viewed as a sign of opinion leadership and may correlate with becoming early members.
    \item {\em Language distance from the community.}
    A user's language distance from the community can relate to becoming an early member in either direction: 
    a small language distance may suggest that the user fits in well and finds it hedonistic to explore, while a large language distance suggests that the user prefers something different and may move to a new community.
    We measure a user's language distance from the community using cross entropy of her posts from the community unigram language model following \citet{Danescu-Niculescu-Mizil:2013:NCO:2488388.2488416,tan2015all}.
    In addition to average distance, we also compute the standard deviation of distances.
\end{itemize}

\begin{figure}[t]
    \centering
    \begin{subfigure}[t]{0.23\textwidth}
        \addFigure{\textwidth}{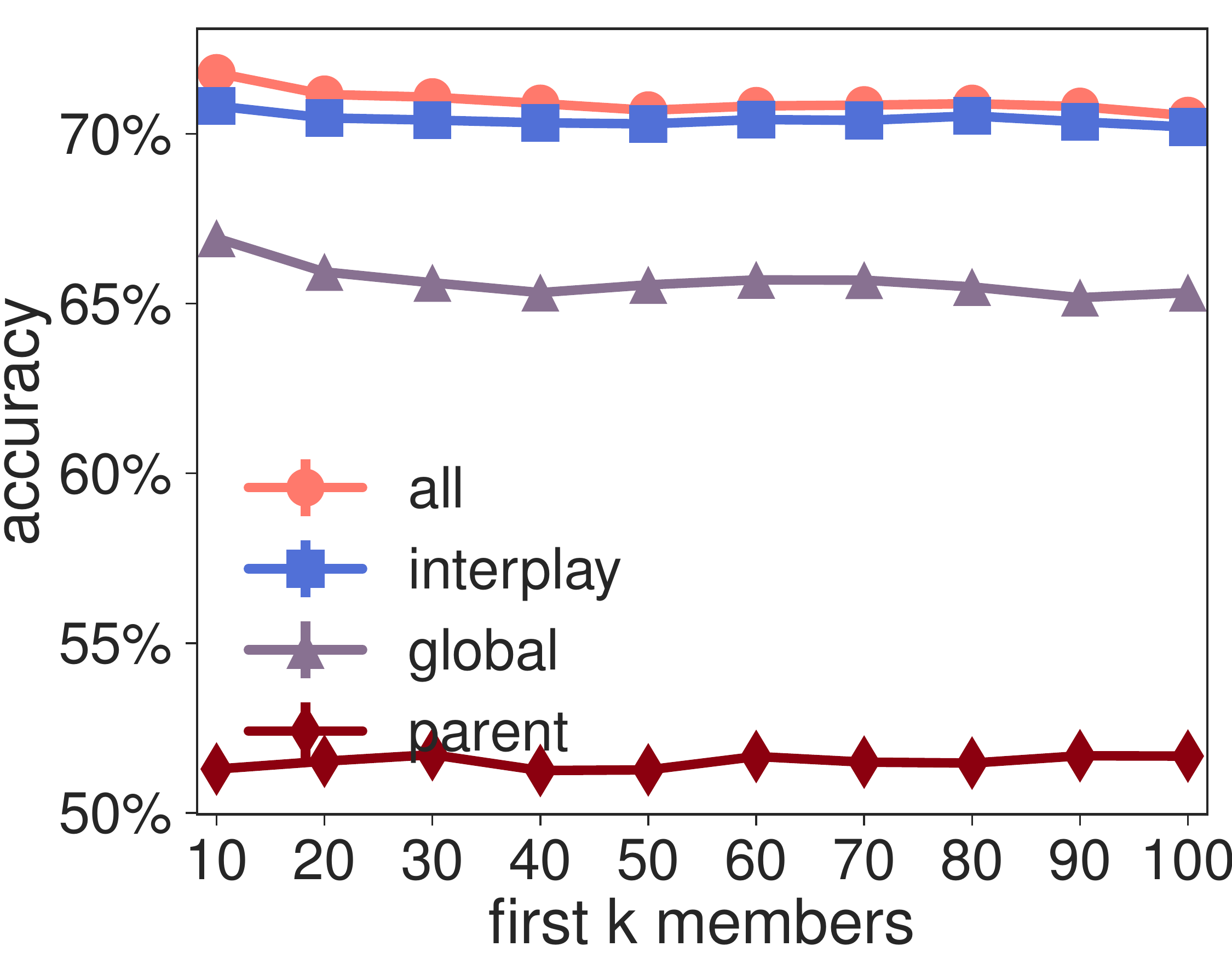}
        \caption{Prediction performance.}
        \label{fig:fp_random_pairs}
    \end{subfigure}
    \hfill
    \begin{subfigure}[t]{0.23\textwidth}
        \addFigure{\textwidth}{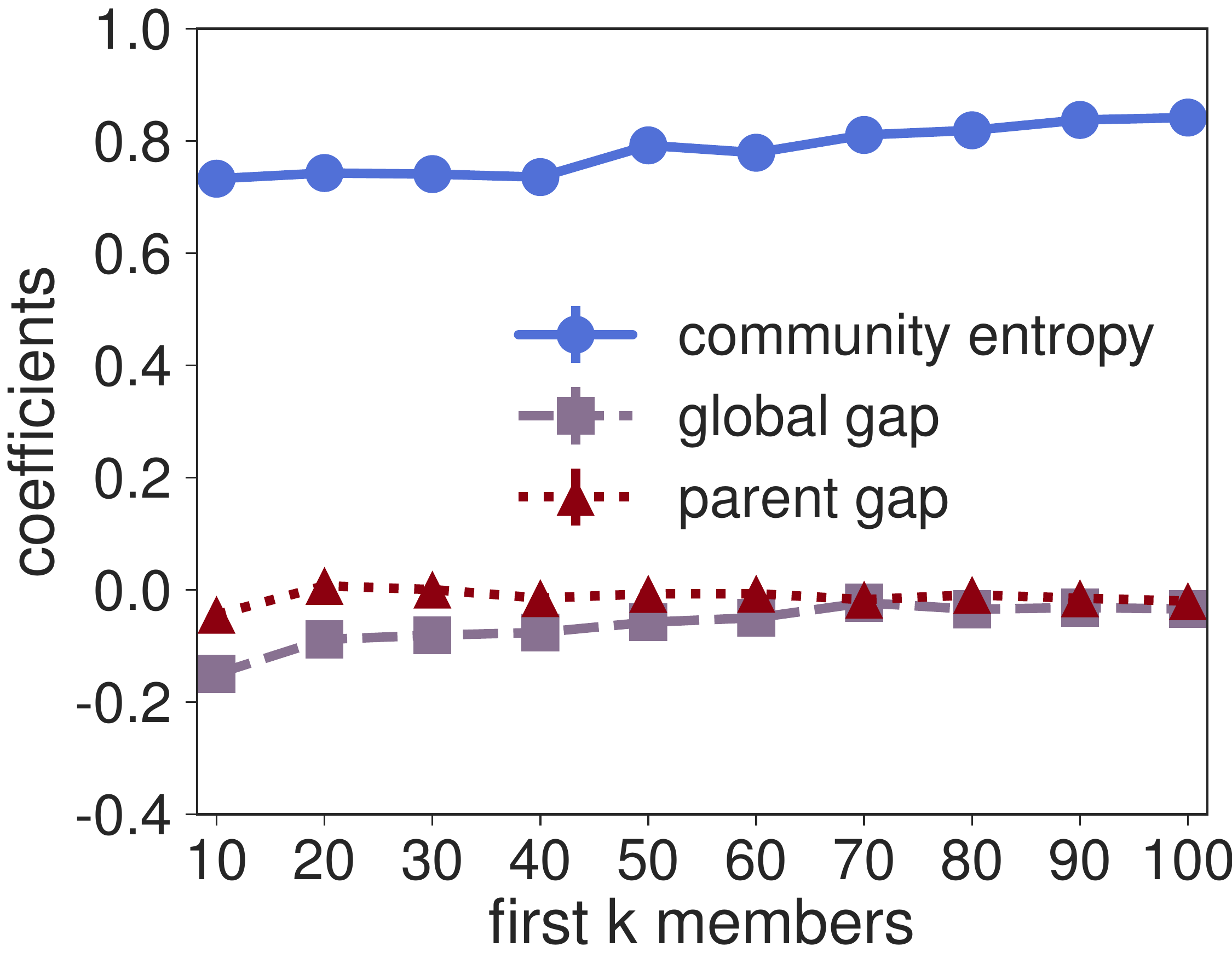}
        \caption{Feature coefficients.}
        \label{fig:fp_coefs}
    \end{subfigure}
    \caption{Predicting whether an existing user is going to become an early member in a child community.
    \figref{fig:fp_random_pairs} shows the prediction accuracy for different feature sets as $k$ increases. Interplay features dominate the performance.
    \figref{fig:fp_coefs} shows coefficients of a feature in each feature set. Community entropy is more important than average time gap in the parent community and in the entire Reddit website. 
    Error bars represent standard errors.
    \label{fig:founder_prediction}}
\end{figure}

We first use hypothesis testing to evaluate single features when $k=100$ (\tableref{tb:founder_prediction}) and then experiment in a prediction framework to examine the effectiveness of these features.

\para{Small time gap in the parent community is negatively correlated with becoming an early member (the {\em parent} column in \tableref{tb:founder_prediction}).}
Different from our hypothesis, 
our observation suggests that controlling for activity levels, users with a short burst of posts in the parent community are less likely to participate in the child community.
Since we match users based on the number of posts in the parent community, it is expected that \#posts in the parent community does not matter.
However, little signal exists in language distance or community feedback in the parent community.

\para{Users that fit well in language globally are more likely to become an early member (the {\em global} column in \tableref{tb:founder_prediction}).}
Results on both the number of posts and average time gap are consistent with our hypothesis that active users are likely to become early members of the child community.
Early members tend to fit well in language use in the entire Reddit website (smaller language distance).
This is different from the characteristic of long-term staying users in \citet{tan2015all} whose language use is more distant from the community than that of early-departing users.
Intriguingly, larger standard deviation in language distance is correlated with becoming an early member, which suggests that diverse language use is %
a sign of early members.
Again, community feedback does not seem to matter, which suggests that early members do not tend to be opinion leaders.

\para{Early members have a diverse portfolio across communities (the {\em interplay} column in \tableref{tb:founder_prediction}).}
We use interplay features to capture a user's behavior in the parent community in the context of the entire Reddit.
We compute the fraction of posts in the parent community and the entropy of posts across existing communities.
We find that a diverse portfolio across existing communities is associated with a large likelihood to become an early member, indicating that early members share properties with market mavens.

\subsection{Prediction Performance}
\label{sec:founder_prediction}

We use logistic regression with $\ell_2$ regularization and the same prediction setup regarding data split, feature normalization, and grid search as in predicting community growth.

\para{Interplay features dominate (\figref{fig:fp_random_pairs}).}
It turns out that although there are only two features in interplay features, they dominate the performance in predicting early members regardless of choices of $k$.
The performance of using all features is relatively {\em stable} across choices of $k$, which suggests that characteristics of early members are insensitive to choices of $k$.
Using all features slightly outperforms using only interplay features when $k=100$ (70.5\% vs. 70.2\%, $p<0.0001$).
Features based on the parent community provide the worst performance, barely above the baseline, indicating that community feedback and language use in the parent community do not provide much signal when we control for user activity.

Feature coefficients (\figref{fig:fp_coefs}) show a similar story that early members of a community are similar to ``market mavens'' in the domain of innovation adoption, who have information about many products and markets \cite{Feick:JournalOfMarketing:1987}. 
The coefficient of community entropy dominates behavioral features based on the parent community alone and the entire Reddit website. The coefficient of average time gap in the parent community, for instance, is very close to 0, which explains the low performance of ``parent'' in \figref{fig:fp_random_pairs}.

Our results show that although our task is set up based on a (parent, child) tuple in the genealogy graph, information regarding the user as a whole is valuable for predicting early members.
Therefore, it is important that we take a holistic view of a user's past history and consider the complete sequence of user behavior if possible.

\section{Concluding Discussion}

In this work, we study the emergence of communities and present the first large-scale characterization of genealogy graphs between communities,
which are defined using previous community memberships of a community's early members.
We show that as a community emerges, the number of its parents increases and parent weights decrease. 
Intriguingly, despite the fact that the number of communities on Reddit increases rapidly over time, the number of parents and max parent weight converge to a stable state rather quickly.
We also demonstrate the effectiveness of using a community's origin to predict its future growth, especially in predicting growth rate.
We find that strong connections with parent communities are related to a community's future growth, which echoes the idea of complex contagion that 
the diffusion of certain behavior requires dense connections between early adopters
\cite{Centola:AmericanJournalOfSociology:}.
Finally, we explore the characteristics of early members and find that a diverse portfolio across communities is a key factor.

Our work constitutes a first step towards understanding the emerging process of a community in the context of existing communities.
Our observations open up future research directions.
For instance, the convergence of basic graph properties over time may relate to the cognitive limit of individuals \cite{Dunbar:JournalOfHumanEvolution:1992} or the nature of building a new community, namely, the emergence of a new community requires a particular structure among early members ($\sim$10\% of early members share experience in an existing community). 
It remains an open question to explain such convergence and investigate whether similar convergence happens on other platforms or based on other membership definitions.

Furthermore, the notion of genealogy graphs has broad implications for understanding communities and requires much more exploration beyond our proposed approach.
Although we focus on a small constant number of early members to examine the emerging process, the emerging process is inherently dynamic and requires varying user bases and time scales across communities.
Future research may develop a robust way to adapt the definition of early members to different communities and even a general method to identify different stages of a community including the emerging process.
Another interesting question is to examine the differences in a community's origin and the characteristics of early members across different types of communities,
e.g., communities on different topics (politics vs. gaming), or common-bond groups vs. common-identity groups \cite{prentice1994asymmetries,YuqingRen:OrganizationStudies:2007}.
Last but not least, each member may not contribute equally in a genealogy graph.
The nature of an edge between a parent community and a child community can further differ depending on why these members join the child community.
It is a promising direction to build genealogy graphs with rich information on edges.

\small
\para{Acknowledgments.} 
We thank J. Hessel, B. Keegan, V. Lai, L. Lee, N. Li, M. Naaman, A. Sharma, S. Zhang, and
anonymous reviewers for helpful comments and discussions.
We thank J. Baumgartner and J. Hessel for sharing the dataset that enabled this research.
This research was supported in part by a gift from Facebook.

\balance
\small
\bibliographystyle{aaai}
\bibliography{reference}

\end{document}